\documentclass[reprint,amsmath,amssymb,aps,prb]{revtex4-2}

\usepackage{graphicx}
\usepackage{amsmath}
\usepackage{xcolor}
\usepackage{dcolumn}% Align table columns on decimal point
\usepackage{enumitem}
\usepackage{nicefrac}

\begin{document}

\title{Antiferromagnetic resonances in superconductor-ferromagnet multilayers}

\author{I.~A.~Golovchanskiy$^{1,2,3,*}$, V.~V.~Ryazanov$^{1,2,4}$, V.~S.~Stolyarov$^{1,2,3}$}

\affiliation{
$^1$ Moscow Institute of Physics and Technology, State University, 9 Institutskiy per., Dolgoprudny, Moscow Region, 141700, Russia; 
$^2$ National University of Science and Technology MISIS, 4 Leninsky prosp., Moscow, 119049, Russia; 
$^3$ Dukhov Research Institute of Automatics (VNIIA), 127055 Moscow, Russia; 
$^4$ Institute of Solid State Physics (ISSP RAS), Chernogolovka, 142432, Moscow region, Russia. 
%$^5$ Faculty of Science and Technology and MESA+ Institute for Nanotechnology, University of Twente, 7500 AE Enschede, The Netherlands.
%$^8$ Physikalisches Institut, Karlsruhe Institute of Technology, 76131 Karlsruhe, Germany; \\
%$^9$ Russian Quantum Center, Skolkovo, Moscow 143025, Russia; \\
%$^{10}$ Skobeltsyn Institute of Nuclear Physics, MSU, Moscow, 119991, Russia
}%

\begin{abstract}
In this work, we study magnetization dynamics in superconductor-ferromagnet (S-F) thin-film multilayer.
Theoretical considerations supported by the broad-band ferromagnetic resonance spectroscopy reveal development of acoustic and optic resonance modes in S-F multilayers at significantly higher frequencies in comparison to the Kittel mode of individual F-layers.
These modes are formed due to antiferromagnetic-like interaction between F-layers via shared circulating superconducting currents in S-layers .
The gap between resonance modes is determined by the thickness and superconducting penetration depth in S-layers.
Overall, rich spectrum of S-F multilayers and its tunability opens wide prospects for application of these multialyers in magnonics as well as in various superconducting hybrid systems.
\end{abstract}

\maketitle

%%%%%%%%%%%%%%%%%%%%%%%%%%%%
%%%%%%%%%%%%%%%%%%%%%%%%%%%%
%%%%%%%%%%%%%%%%%%%%%%%%%%%%
%\section{Introduction}
{\it Introduction.}
Hybridization of antagonistic superconducting (S) and ferromagnetic (F) orders offers in electronics and spintronics, which have been repeatedly demonstrated in past decades \cite{Linder_NatPhys_11_307}. 
%The interplay between ferromagnetic and superconducting spin orders enables manipulation with spin states and leads to a development of various electronic and spintronic elements, including  superconductor-ferromagnet-superconductor (S-F-S) Josephson junctions \cite{Linder_NatPhys_11_307,Ryazanov_PRL_86_2427,Weides_APL_89_122511}, superconducting phase shifters \cite{Yamashita_SRep_10_13687,Feofanov_NatPhys_6_593}, memory elements \cite{Golovchanskiy_PRB_94_214514,Vernik_IEEETAS_23_1701208,Karelina_JAP_130_173901}, F-S-F–based spin valves \cite{Gingrich_NatPhys_12_564,Lenk_PRB_96_184521}, Josephson diodes \cite{Jeon_NatMat_2022} and more complex long-range spin-triplet superconducting systems \cite{Robinson_Sci_329_59,Banerjee_NatComm_5_4771,Wang_NatPhys_6_389,Glick_SciAdv_4_eaat9457,Jeon_NatMat_20_1358}.
Recently the interest in S-F hybridization has been reinforced by demonstrations of its prospects in relation with the magnetization dynamics phenomena. 
%Magnonics is a growing field of research which offers approaches for the transfer and processing of information via spin waves.
%A good overview of various potential applications and recent advances in magnonics can be found in Refs.~\cite{Chumak_IEEE_58_0800172,Pirro_NatRevMat_6_1114,Barman_JPCM_33_413001,Chumak_NatPhys_11_453,Lenk_PhysRep_507_107,Csaba_PLA_381_1471,Serga_JPDAP_43_264002,Kruglyak_JPDAP_43_264001} and references therein.
%In development of magnonic systems one of principle requirements is engineering of appropriate spin-wave dispersion.  
%Various wide-range manipulations with the spin-wave dispersion have been demonstrated at cryogenic temperatures when coupling magnonic systems with superconductors.
In particular, interactions between magnetization dynamics and the superconducting vortex lattice allow to form and guide the tunable magnonic band structure \cite{Dobrovolskiy_NatPhys_15_477}, as well as to induce exchange spin waves by the DC electric current \cite{Dobrovolskiy_arXiv_2103.10156}. 
Also, interactions between magnetization dynamics and superconducting Meissner currents in hybrid structures modifies the spin-wave dispersion \cite{Golovchanskiy_AFM_28_1802375,Golovchanskiy_JAP_124_233903}, which can be used for creation of magnonic crystals \cite{Golovchanskiy_AdvSci_6_1900435} or for gating magnon currents \cite{Yu_2201_09532}.
Remarkably, low speed of electromagnetic propagation in superconductor-insulator-superconductor thin-film structures facilitates achievement of the ultra-strong photon-to-magnon coupling in on-chip hybrid devices \cite{Golovchanskiy_SciAdv_7_eabe8638,Golovchanskiy_PRAppl_16_034029} aiming for the photon-to-magnon entanglement\cite{Silaev_arxiv2}. 
 
A new strong phenomenon in S-F hybrid structures was reported recently in Refs.~\cite{Li_ChPL_35_077401,Jeon_PRAppl_11_014061,Golovchanskiy_PRAppl_14_024086} and investigated further in Refs.~\cite{Golovchanskiy_arxiv,Silaev}. 
In superconductor-ferromagnet-superconductor (S-F-S) thin-film structures in the presence of electronic interaction between superconducting and ferromagnetic layers a radical increase in the ferromagnetic resonance (FMR) frequency occurs.
The mechanism behind the phenomenon constitutes a formation of one-dimensional superconducting torque via the interplay between the superconducting imaginary conductance and magnetization precession at S-F interfaces, which result in induction of alternating circulating superconducting currents in the opposite phase to the magnetization precession.

In this work, we generalize the problem and consider magnetization dynamics in arbitrary S-F multilayers. 
Coupling between ferromagnetic layers via superconducting currents allows to induce antiferromagnetic-like interaction between F-layers, which result in acoustic and optic resonances modes.
Theoretical considerations supported by the broadband ferromagnetic resonance spectroscopy demonstrate that the spectrum is determined by geometrical characteristics of a multilayer as well as by the superconducting penetration depth in S-layers.

\begin{figure}[!ht]
\begin{center}
\includegraphics[width=0.7\columnwidth]{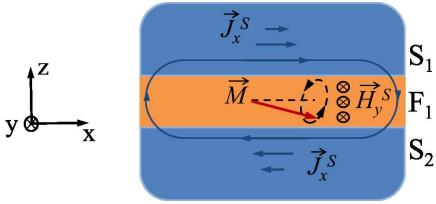}
\caption{
Schematic illustration of the interplay between ac magnetic field, magnetization precession and superconducting currents in S-F-S trilayer.
Magnetization precession ($\vec{M}$, black arrow) at S-F interfaces induces macroscopic
superconducting currents alternating in S-layers along the $x$-direction ($J^S_x$, blue arrows). 
These currents form the magnetic field $H^S_y$ in the F-layer along the $y$-direction in opposite phase to the precession of $\vec{M}$.
}
\label{Fig1}
\end{center}
\end{figure}

%%%%%%%%%%%%%%%%%%%%%%%%%%%%
%%%%%%%%%%%%%%%%%%%%%%%%%%%%
%%%%%%%%%%%%%%%%%%%%%%%%%%%%
%\section{Theory}
{\it Theory.}
Following Refs.~\cite{Silaev,Kostylev_JAP_106_043903}, electrodynamics and magnetization dynamics in S-F multilayers obeys conventional Maxwell equations supplemented by the Ohm law with imaginary conductance in superconducting layers and by the Polder susceptibility in ferromagnetic layers.
By neglecting edge effects, the $y$-component of magnetic field as well as $x$-components of electric field and of the current are functions of the transverse coordinate $z$ only (see Fig.~\ref{Fig1}).
Derivation of the magnetic field in S- and F-layers from initial Maxwell equations yields following general expressions
\begin{equation}
\begin{aligned}
H^S_y(z) & = A_i\exp{\frac{z}{\lambda_S}}+B_i\exp{-\frac{z}{\lambda_S}}, \\
H^F_y(z) & = C_i\exp{\frac{z}{\lambda_F}}+D_i\exp{-\frac{z}{\lambda_F}},
\label{inSF}
\end{aligned}
\end{equation}
where the subscript of coefficients $i$ specifies the superconducting layer or the ferromagnetic layer in the stack, $\lambda_S$ is the superconducting penetration depth, and $\lambda_F=\delta_F\Omega$ is the ferromagnetic penetration depth, $\delta_F=\sqrt{i/\mu_0\omega\sigma_F}$ is the conventional electromagnetic penetration depth into a metal with conductivity $\sigma_F$ (typically, in permalloy $\sigma_F\sim10^6$~Ohm$^{-1}$m$^{-1}$).
The characteristic dimensionless frequency $\Omega$ is given by
\begin{equation}
\Omega^2=\frac{\gamma^2(H+H_a)(H+H_a+M_{eff})-\omega^2}{\gamma^2(H+H_a+M_{eff})^2-\omega^2},
\label{Om}
\end{equation}
where $H$ is the external field (aligned with the $x$-axis in Fig.~\ref{Fig1}), $H_a$ is the effective field of the uniaxial anisotropy aligned with the external field, and $M_{eff}=M_s-2K_u/\mu_0M_s$ is the effective magnetization, which accounts the out-of-plane uniaxial anisotropy with the constant $K_u$. 
Notice that the conventional Kittel formula for the ferromagnetic resonance in thin films in these notations is provided by $\Omega=0$.
At every S/F interface the following boundary conditions are fulfilled
\begin{equation}
\begin{aligned}
H^S_y &= H^F_y, \\
\frac{1}{\sigma_S}\frac{dH^S_y}{dz} &= \frac{1}{\sigma_F}\frac{dH^F_y}{dz},
\label{BC}
\end{aligned}
\end{equation}
where $\sigma_S=i/\mu_0\omega\lambda_S^2$ is the imaginary conductance in superconducting layers.

\begin{figure}[!ht]
\begin{center}
\includegraphics[width=1\columnwidth]{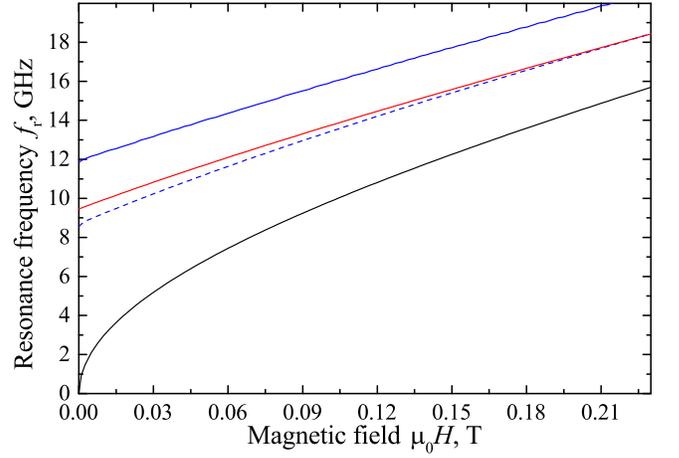}
\caption{
Theoretical dependencies of the resonance frequency on the magnetic field $f_r(H)$ in conventional F-layer (solid black curve, Kittel formula), S-F-S trilayer (solid red curve, Eq.~\ref{SFS}) and symmetric S-F-S-F-S multilayer (solid and dashed blue curves, Eq.~\ref{SFSFS}).
The following parameters are used for calculations: $\gamma/2\pi=29.5$~GHz/T, $d_{F1}=d_{F2}=50$~nm, $d_{S1}=d_{S3}=d_{Se}=150$~nm, $d_{S2}=d_{Si}=100$~nm, $\lambda_S=100$~nm, $\mu_0M_{eff}=1$~T, $H_a=0$.
}
\label{th}
\end{center}
\end{figure}

For the S-F-S trilayer, depicted schematically in Fig.~\ref{Fig1}, solution of Eq.\ref{BC} together with natural boundary conditions at outer surfaces of S-layers, namely, $H^S_y=0$, and in the limit $d_F\ll \lambda_F$ yields the following expression for ferromagnetic resonance frequency:
\begin{equation}
\Omega^2=-\frac{d_F}{\lambda_S}\frac{\tanh{d_{S1}/\lambda_S}\tanh{d_{S2}/\lambda_S}}{\tanh{d_{S1}/\lambda_S}+\tanh{d_{S2}/\lambda_S}}.
\label{SFS}
\end{equation}
As example, black and red curves in Fig.~\ref{th} compares the dependence of the resonance frequency on the magnetic field $f_r(H)$, respectively, in a single F-layer and in S-F-S trilayer with the following thicknesses: $d_{F1}=50$~nm, $d_{S1}=150$~nm, $d_{S2}=100$~nm, $\lambda_S=100$~nm.

\begin{figure}[!ht]
\begin{center}
\includegraphics[width=1\columnwidth]{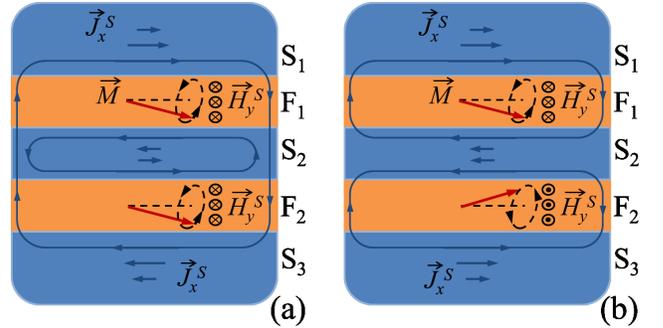}
\caption{
Schematic illustration of the interplay between ac magnetic field, magnetization precession and superconducting currents in symmetric S-F-S-F-S multilayer.
Acoustic (a) and optic (b) modes a formed. 
}
\label{Fig2}
\end{center}
\end{figure}

Application of the same derivation approach for the symmetric S-F-S-F-S multilayer, depicted schematically in Fig.~\ref{Fig2}, yields two resonance modes with the in-phase (acoustic mode) and the anti-phase (optic mode) precession of ferromagnetic layers, respectively,
\begin{equation}
\begin{aligned}
\Omega^2_a & = -\frac{d_F}{\lambda_S}\frac{\tanh{d_{Si}/\lambda_S}\coth{d_{Se}/2\lambda_S}}{\tanh{d_{Si}/\lambda_S}+\coth{d_{Se}/2\lambda_S}}, \\
\Omega^2_o & = -\frac{d_F}{\lambda_S}\frac{\tanh{d_{Si}/\lambda_S}\tanh{d_{Se}/2\lambda_S}}{\tanh{d_{Si}/\lambda_S}+\tanh{d_{Se}/2\lambda_S}}, 
\label{SFSFS}
\end{aligned}
\end{equation}
where $Se$ denotes external superconducting layers (S1 and S3 in Fig.~\ref{Fig2}), $Si$ corresponds to the internal superconducting layer (S2 in Fig.~\ref{Fig2}), and thicknesses of both ferromagnetic layers is considered equal $d_{F1}=d_{F2}=d_F$.
Blue curves in Fig.~\ref{th} show acoustic and optic resonance curves $f_r(H)$ in the multilayer with the same thicknesses as in a single F-layer and in S-F-S trilayer : $d_{F1}=d_{F2}=d_F=50$~nm, $d_{S1}=d_{S3}=d_{Se}=150$~nm, $d_{Si}=100$~nm, $\lambda_S=100$~nm.

\begin{figure*}[!ht]
\begin{center}
\includegraphics[width=0.66\columnwidth]{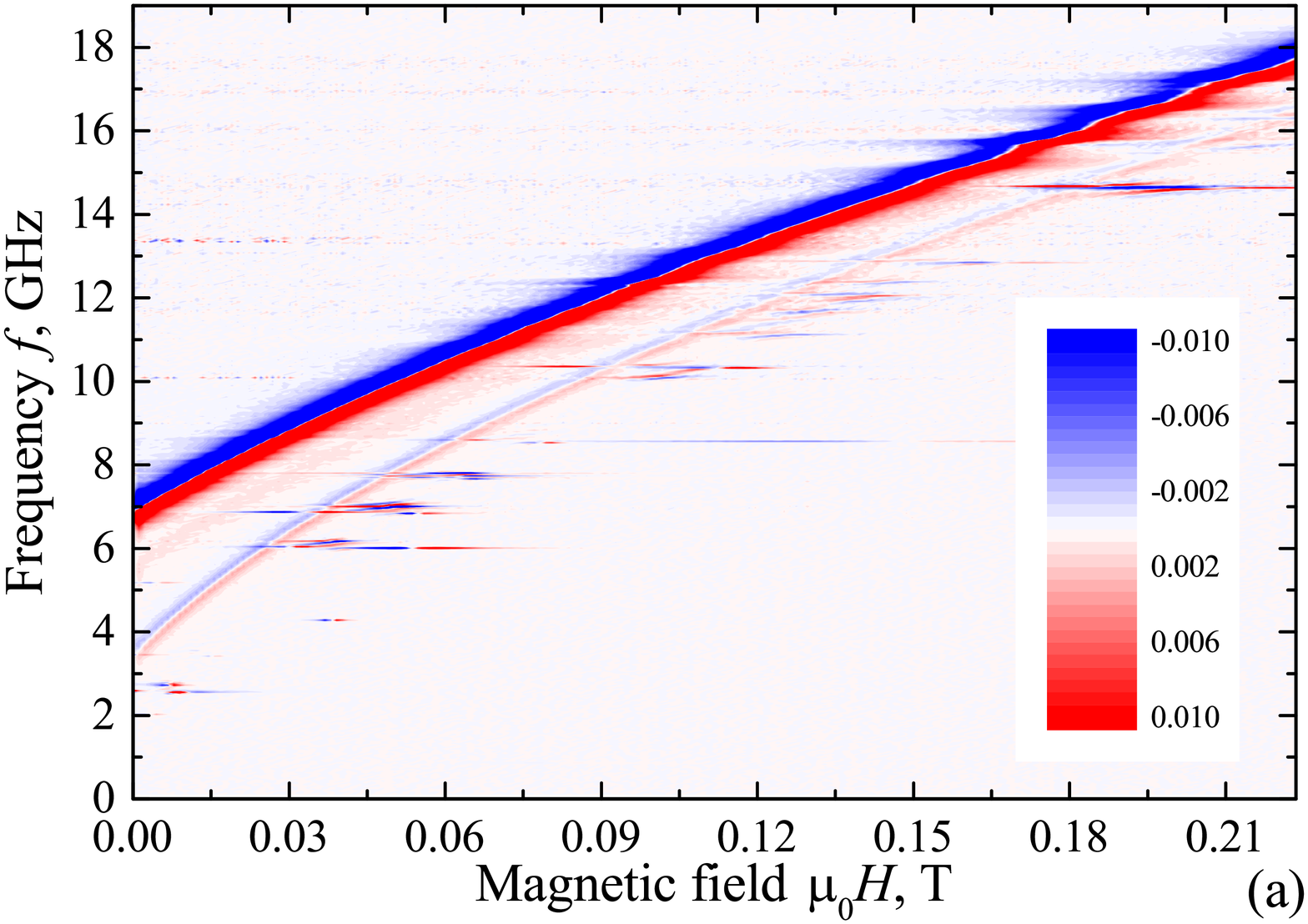}
\includegraphics[width=0.66\columnwidth]{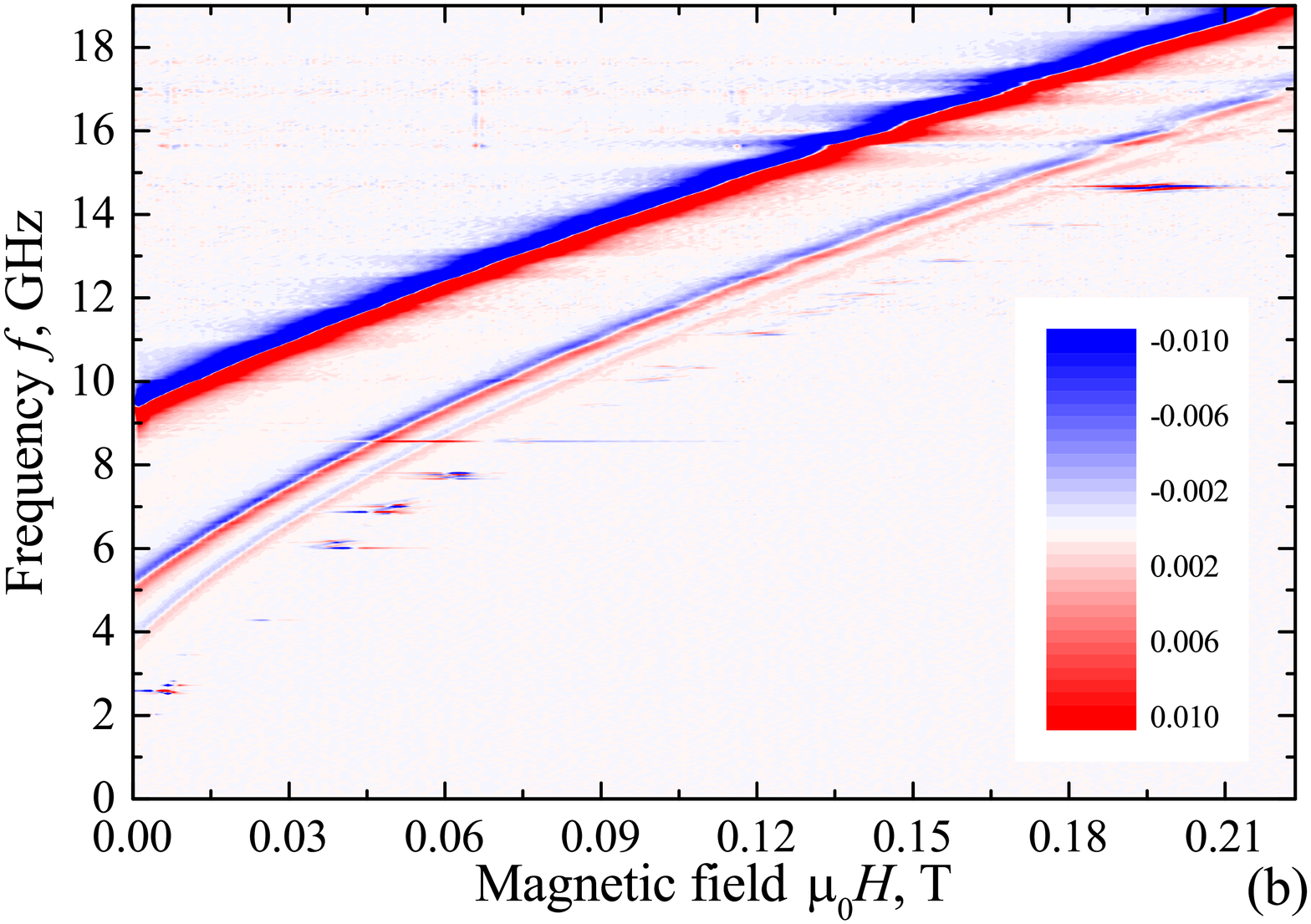}
\includegraphics[width=0.66\columnwidth]{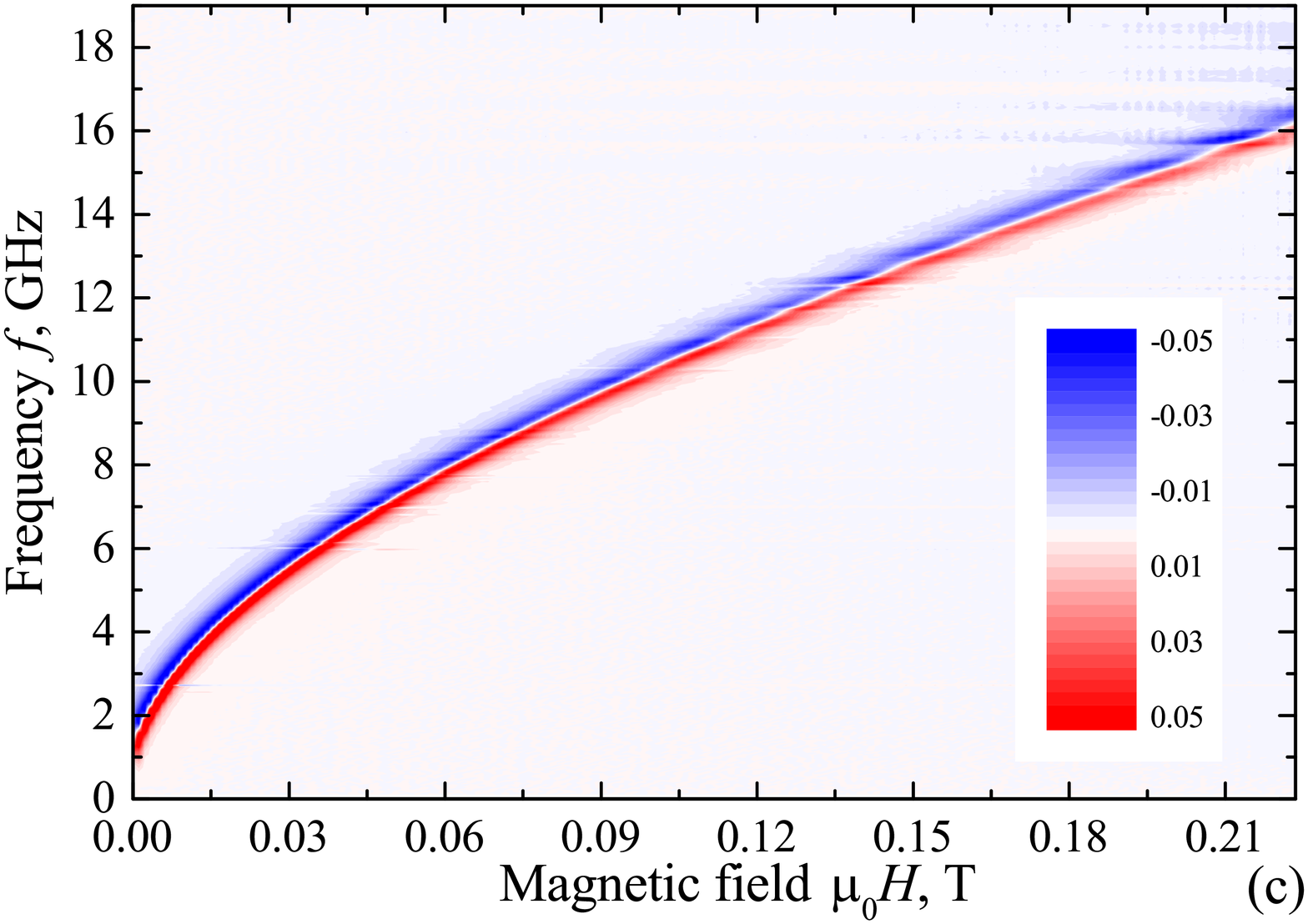}
\includegraphics[width=0.66\columnwidth]{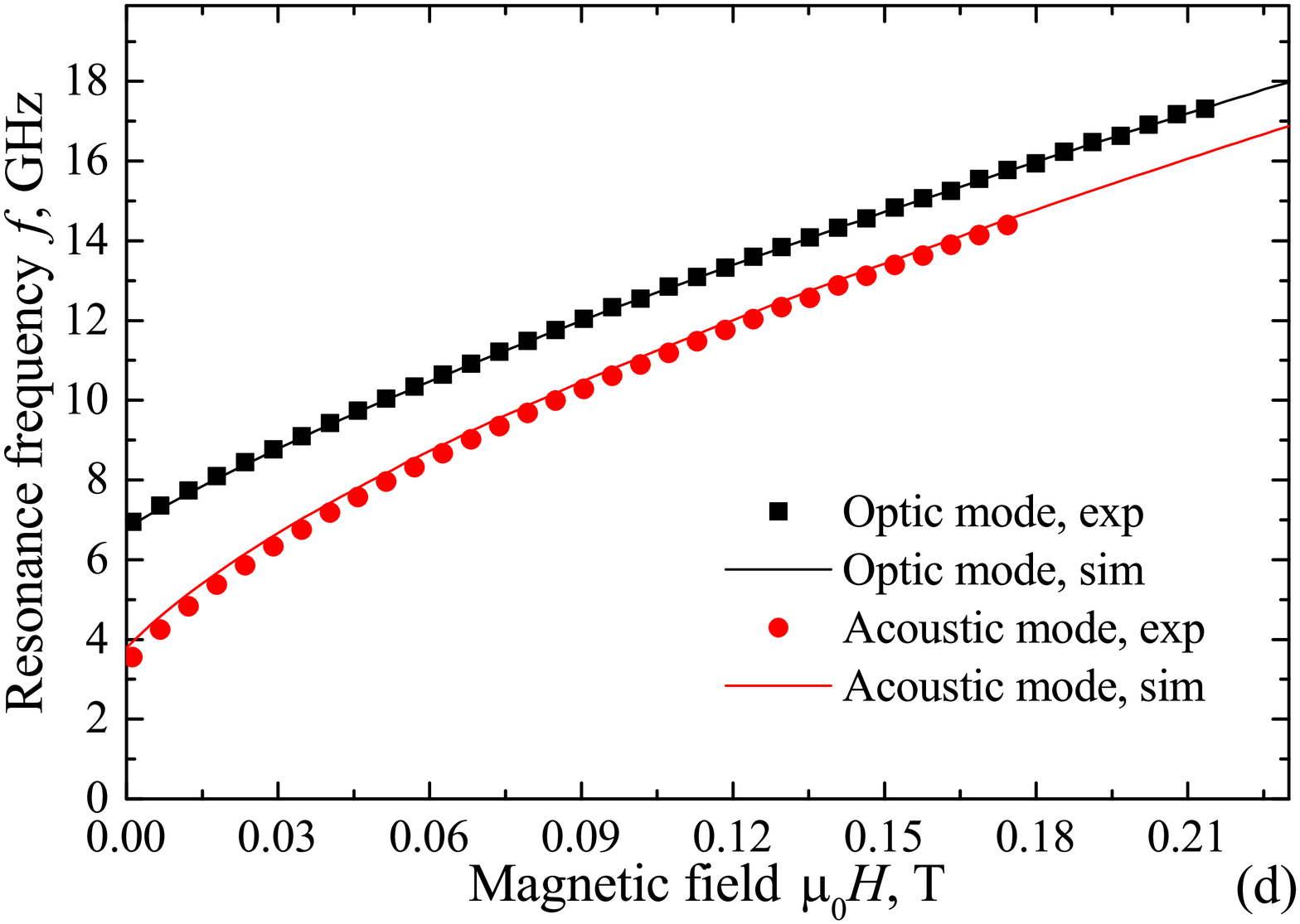}
\includegraphics[width=0.66\columnwidth]{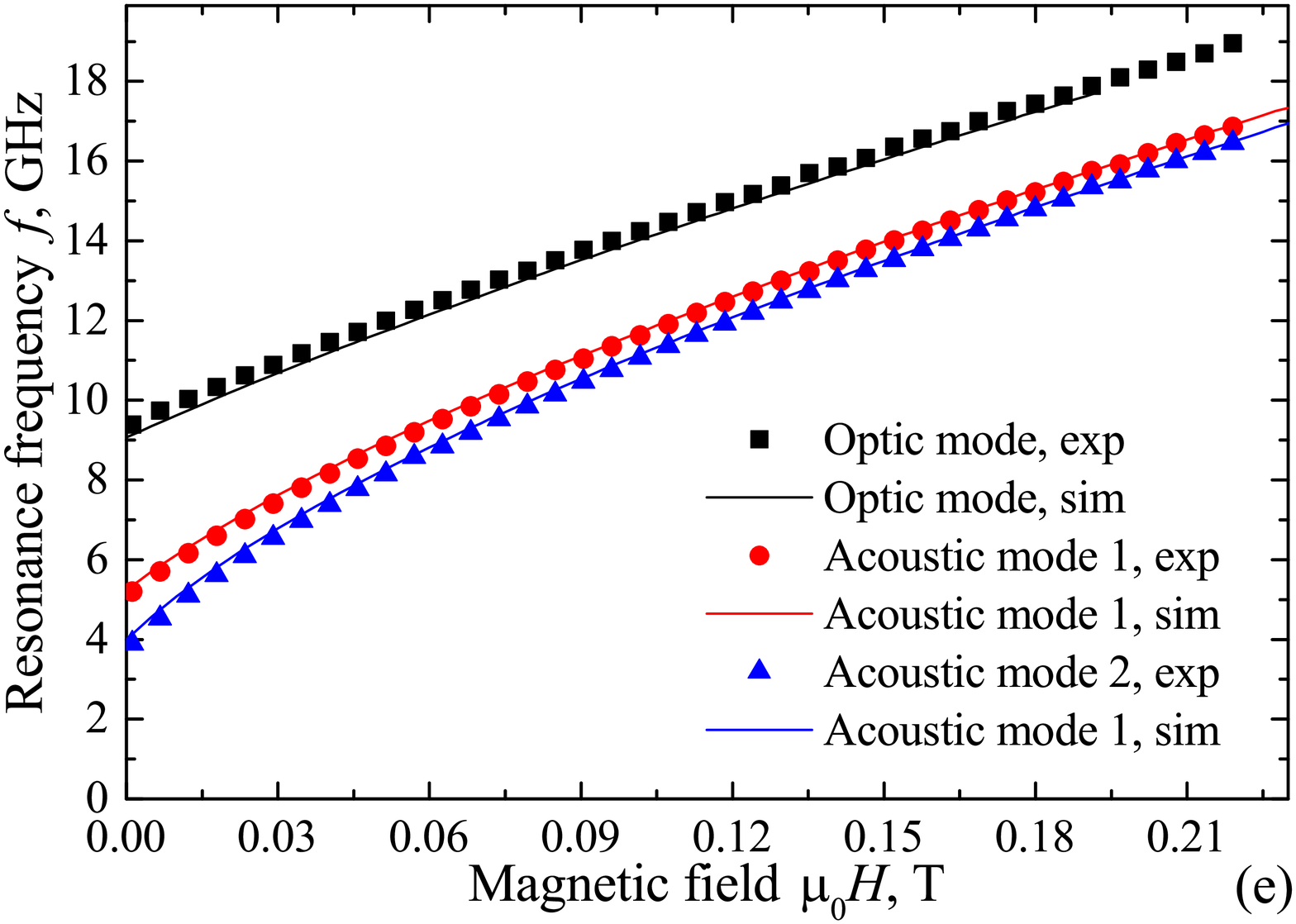}
\caption{
a-c) Differentiated transmission spectra $dS_{21}/dH(f,H)$ for samples SF1 (a) and SF2 (b) at temperature 2 K and for sample SF1 at 8~K (c). 
Colour codes are provided in insets.
d,f) Experimental (symbols) and theoretical (solid curves) resonance lines for samples SF1 (a) and SF2 (b) at temperature 2 K.
}
\label{Exp}
\end{center}
\end{figure*}

It should be noticed that the acoustic mode (Fig.~\ref{Fig2}a) can be thought as being formed by the global superconducting current, which circulates in external superconducting layers S1 and S3, while the internal S-layer S2 only screens the induced magnetic field in conventional manner.  
In fact, in the limit $d_{Se}\rightarrow\infty$ the expression for $\Omega_a$ in Eq.~\ref{SFSFS} meets $\Omega$ in Eq.\ref{SFS} with the substitution $d_F\rightarrow 2d_F$.
The optic mode (Fig.~\ref{Fig2}b) can be thought as the resonance in S-F-S trilayer with reduced thickness of internal S-layer: the expression for $\Omega_o$ in Eq.~\ref{SFSFS} meets $\Omega$ in Eq.~\ref{SFS} with the substitution $d_{S2}=d_{Si}/2$.
Qualitatively it can be concluded that the optic mode is unaffected by the interaction between F-layers layers, while the acoustic mode gains energy due to the coupling.
Interestingly, this qualitative picture is in direct contradiction with magnetization dynamics in exchange-coupled ferromagnetic layers \cite{Schmool_JPCM_10_10679,Lindner_JPCM_15_R193, Rezende_JAP_126_151101, Golovchanskiy_JAP_131_053901, Golovchanskiy_PRB_106_024412}, where regardless the details of the exchange interaction the acoustic mode corresponds to magnetization dynamics in non-interacting magnetic layers.
Also, according to Fig.~\ref{th} and Eq.~\ref{SFSFS} the optic mode is observed at lower frequencies in comparison to the acoustic mode, which characterises the coupling between ferromagnetic layers via superconducting layers as antiferromagnetic.
In a way, such interaction between ferromagnetic layers via superconducting currents in adjacent layers reminds interaction of fluxons in superconductor-insulator Josephson junction stacks \cite{Holst_PRB_42_127,Sakaki_JAP_73_2411,Ustinov_PRB_54_6111}.

In the general case, resonance modes of an arbitrary S-F multilayer, which consist of $N$ ferromagnetic layers and $N+1$ superconducting layers, can be derived numerically from the set of equations~\ref{BC} in the matrix form, $[M]\times[A_n,B_n,C_n,D_n]^T=0$, by finding frequencies $\omega_r$ that obey the expression
\begin{equation}
\det[M(\omega_r)]=0.
\label{det}
\end{equation}
%

%%%%%%%%%%%%%%%%%%%%%%%%%%%%
%%%%%%%%%%%%%%%%%%%%%%%%%%%%
%%%%%%%%%%%%%%%%%%%%%%%%%%%%
%\section{Experimental details and results}
%\label{fab}
{\it Experimental details and results.}
Experimentally magnetization dynamics in S-F multilayers is studied by measuring the ferromagnetic resonance absorption spectrum using the VNA-FMR approach \cite{Neudecker_JMMM_307_148,Kalarickal_JAP_99_093909,Chen_JAP_101_09C104} and the same chip layout and experimental setup as in Refs~\cite{Golovchanskiy_PRAppl_14_024086,Golovchanskiy_arxiv}.
A series of niobium-permalloy(Py=Fe$_{20}$Ni$_{80}$)-niobium (Nb-Py) multilayered structures are placed directly on top of the central transmission line of superconducting Nb waveguide.
Deposition of Nb-Py multilayers is performed in a single vacuum cycle ensuring the electron transparency at Nb-Py interfaces. 
Thin Si or AlO$_x$ spacing layer is deposited between Nb co-planar waveguide and the multialyers in order to ensure electrical insulation of the studied samples from the waveguide.
Two test samples have been studied: 
a sample with two ferromagnetic layers, that consist of Nb-101nm/Py-11nm/Nb-41nm/Py-11nm/Nb-41nm, referred to as SF2 and 
a sample with three ferromagnetic layers, that consist of Nb-101nm/Py-11nm/Nb-40nm/Py-11nm/Nb-40nm/Py-12nm/Nb-41nm, referred to as SF3 .
The SF2 sample is made asymmetric on purpose in order to provide a finite dynamic susceptibility of the optic mode, which otherwise is zero and, thus, does not couple to uniform microwave magnetic field of the transmission line.

Microwave spectroscopy of samples was performed by measuring the transmission characteristics $S_{21}(f,H)$ in the closed-cycle cryostat Oxford Instruments Triton (base temperature 1.2 K) equipped with the home-made superconducting solenoid.
Spectroscopy was performed in the field range from -0.22 T to 0.22 T, in the frequency range from 0 up to 20 GHz, and in the temperature range from 2 to 11 K. 
Magnetic field was applied in-plane along the direction of the waveguide (see Ref.~\cite{Golovchanskiy_arxiv}).
FMR spectra at different temperatures were analysed by fitting $S_{21}(f)$ characteristics at specified $H$ and $T$ with the Lorentz curve and, thus, obtaining the dependencies of the resonance frequency on magnetic field $f_r(H)$.

%%%%%%%%%%%%%%%%%%%%%%%%%%%%
%%%%%%%%%%%%%%%%%%%%%%%%%%%%
%%%%%%%%%%%%%%%%%%%%%%%%%%%%
%\section{Experimental results}

Figure~\ref{Exp}a-c demonstrates the essence of the studied phenomenon:
at temperatures below the critical temperature of Nb, $T<T_c$, the transmission spectrum for SF2 sample consist of two spectral lines (Fig.~\ref{Exp}a) and for SF3 sample consist of three spectral lines (Fig.~\ref{Exp}b).
At $T>T_c$ (Fig.~\ref{Exp}c) FMR spectrum for both samples is reduced to a single spectral line, which obeys the conventional Kittel formula (Eq.~\ref{Om}, $\Omega=0$).
For both samples the fit of FMR curves at $T>T_c$ yields negligible anisotropy field $\mu_0H_a\approx 2$~mT, the effective magnetization $\mu_0M_{eff}\approx1.108$~T, which is close to typical values of the saturation magnetization of permalloy $\mu_0 M_s\approx1$~T, and no noticeable dependence of $H_a$ and $M_{eff}$ on temperature.
Temperature dependencies of FMR spectra for both samples yield superconducting critical temperatures $T_c=7.7$~K for SF2 sample and $T_c=7.9$~K for SF3 sample.
The critical temperature of Nb layers is reduced in comparison to the bulk critical temperature of Nb $T_c\approx9$~K owing to the inverse proximity effect \cite{Aarts_PRB_56_2779}.

At $T<T_c$ (Fig.~\ref{Exp}a,b) FMR spectrum shifts to higher frequencies and splits to spectral lines in accordance to the number of F-layers in the stack.
The strongest line, the acoustic mode, is observed at the highest frequencies, while weaker lines at lower frequencies correspond to optic modes.
Resonance lines were modelled with Eq.~\ref{det} using $\lambda_S$ as the fitting parameter (see Fig.~\ref{Exp}d,e). 
The optimum fit is obtained with $\lambda_S=115$~nm for SF2 sample and $\lambda_S=98$~nm for SF3 sample.
The obtained $\lambda_S$ is slightly higher than typical values in bulk Nb (about 80 nm) due to the inverse proximity effect \cite{Aarts_PRB_56_2779,Silaev}.
A better fit could be obtained by considering a variation of $\lambda_S$ in different S-layers.
Thus, the provided theoretical description of the magnetization dynamics phenomenon in arbitrary S-F multilayers is verified.

%%%%%%%%%%%%%%%%%%%%%%%%%%%%
%%%%%%%%%%%%%%%%%%%%%%%%%%%%
%%%%%%%%%%%%%%%%%%%%%%%%%%%%
%\section{Conclusion}
{\it Conclusion.}
Summarising, we report a study of magnetization dynamics in S-F multialyers. 
Theoretical considerations supported by experiments in a wide frequency, field, and temperature ranges show that the coupling between ferromagnetic layers via superconducting layers results in formation of antiferromagnetic interaction between F-layers with the strength that depends of thickness and superconducting properties of S-layers.
This interaction between ferromagnetic layers is formed via superconducting currents and result in formation of acoustic and optic spectral branches.
These results open wide prospects for application of S-F multialyers in magnonics and also bridges magnetization dynamics phenomena with various superconducting circuits \cite{Barnes_SUST_24_024020,Mai_PRB_84_144519,Golovchanskiy_SUST_30_054005}, hybrid devices \cite{Golovchanskiy_SciAdv_7_eabe8638,Golovchanskiy_PRAppl_16_034029}, and metamaterials \cite{Pimenov_PRL_95_247009}.
Moreover, resonance properties of S-F multilayers by changing the superconducting state of S-layers optically\cite{Silaev,Veshchunov_NatComm_7_12801,Magrini_APL_114_142601} or via electric currents.

%%%%%%%%%%%%%%%%%%%%%%%%%%%%
%\section{Acknowledgments}
{\it Acknowledgements}
The authors acknowledge Dr M. Silaev for fruitful discussions.
The research study was financially supported by the Russian Science Foundation (grant N 22-22-00314). 
%The work was financially supported by the Federal Academic Leadership Program "Priority-2030" (NUST MISIS grant No. K2-2022-029)

%\bibliographystyle{apsrev}
\bibliography{A_Bib_SFS_AFMR}

\end{document}